\documentclass[twocolumn,aps]{revtex4-1} 
\usepackage{amsmath} 

\usepackage{graphicx} 

\newcommand{\op}{\boldsymbol}
\begin{document}

\title{Einstein's Recoiling Slit Experiment, Complementarity and Uncertainty}
\author{Tabish Qureshi}
\email{tabish@ctp-jamia.res.in}
\affiliation{Centre for Theoretical Physics, Jamia Millia Islamia,
New Delhi - 110025, India.}
\author{Radhika Vathsan}
\email{radhika@goa.bits-pilani.ac.in}
\affiliation{Department of Physics, BITS Pilani K K Birla Goa Campus, 
Zuarinagar, Goa 403726, India.}
\date{\today}

\begin{abstract}
We analyze Einstein's recoiling slit experiment and point out that the
inevitable entanglement between the particle and the recoiling-slit was
not part of Bohr's reply. We show that if this entanglement is taken
into account, one can provided a simpler answer to Einstein. We also
derive the Englert-Greenberger-Yasin duality relation  from
this entanglement. In addition, we show that the Englert-Greenberger-Yasin
duality relation can also be thought of as a consequence of the {\em sum
uncertainty relation} for certain observables of the recoiling slit.
Thus, the uncertainty relations and entanglement are both an integral part
of the which-way detection process.
\end{abstract}

\maketitle

\section{Introduction}

The two-slit experiment carried out with particles is a testbed of various
foundational ideas in quantum theory. It has been used to exemplify
wave-particle duality and Bohr's complementarity principle \cite{bohr}.
The two-slit experiment captures the essence of quantum theory in such
a fundamental way that Feynman went to the extent of stating that it is a
phenomenon ``which has in it the heart of quantum mechanics; in reality it
contains the {\em only} mystery" of the theory \cite{feynman}.

\begin{figure}[h!]
\centering
\includegraphics[width=3.5 in]{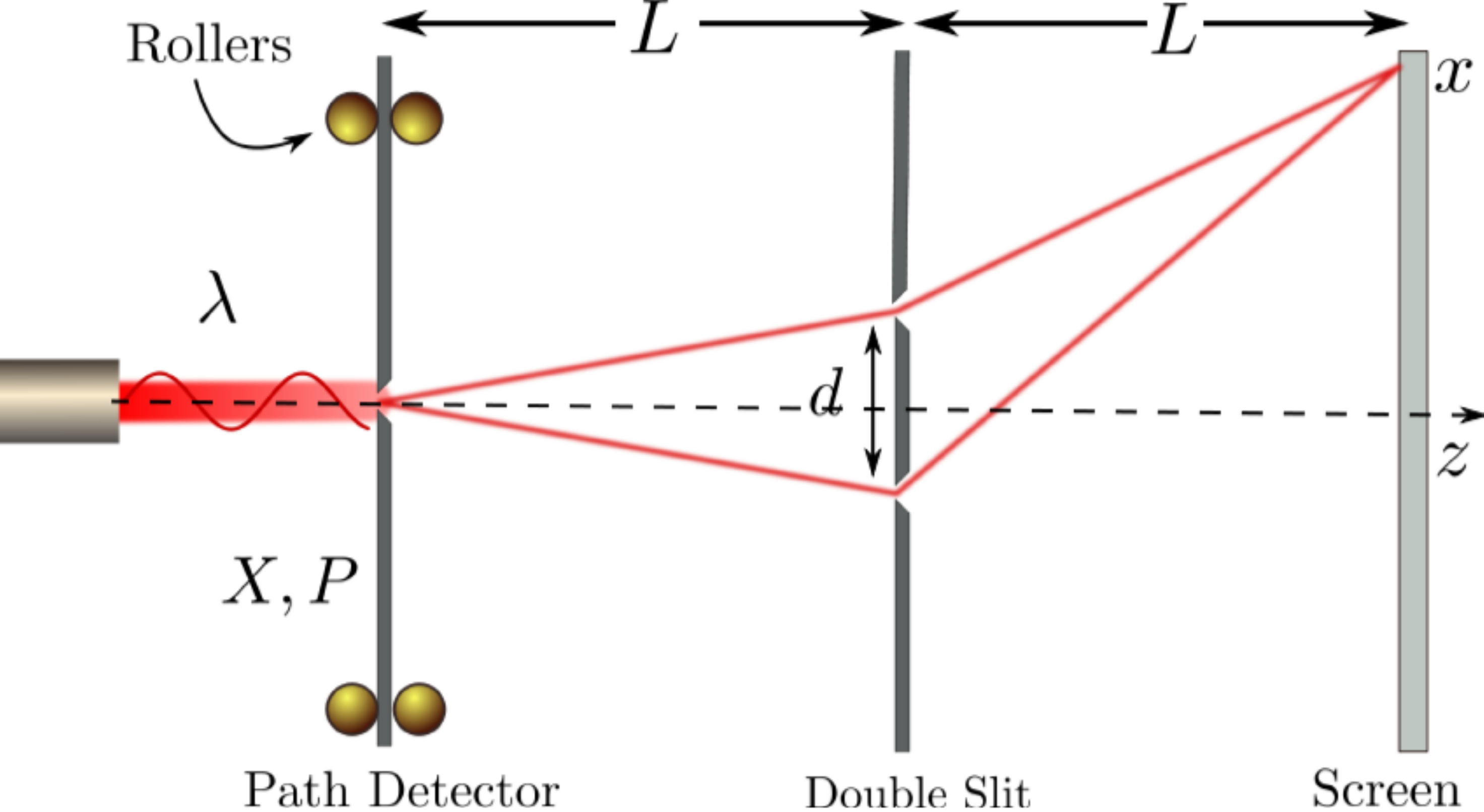}
\caption{Schematic diagram to illustrate Einstein's recoiling slit
experiment.}
\label{recoil}
\end{figure}

Neils Bohr had stressed that the wave-nature of particles, characterized
by interference, and the particle-nature, characterized by the knowledge
of which slit the particle passed through, are mutually exclusive. He
argued that in a single experiment, one could see only one of these two 
complementary properties at a time. Bohr elevated this concept to the
status of a separate principle, the principle of complementarity \cite{bohr}.

Einstein was uncomfortable with the quantum indeterminism and sought to
demonstrate that the principle of indeterminacy could be violated.
Einstein's ongoing criticism of Bohr's view of quantum theory was brought
into focus at the fifth Solvay conference in Brussels in 1927. Einstein
pointed out how it was possible to
use the laws of conservation of energy and momentum to obtain
information on the state of a particle in a process of interference which,
according to the principle of complementarity, should not be accessible.
In the following, we describe Einstein's proposed thought experiment
\cite{recoil}.

\section{Recoiling Slit Experiment}

Light traveling along the $z$-axis, perpendicular to the $x$-axis, is incident on
screen 1 (see FIG. \ref{recoil}) which has a slit narrower than the
wavelength of light. 
The light passes through the single slit and illuminates screen 2 which has two slits.
Light emerging from the double-slit results in the formation of an
interference pattern on the final screen 3.

Einstein suggested that screen 1 be free to slide along
the $x$-axis. According to his argument, the deflection of the light from the original direction of propagation can only be caused by
its interaction with this screen. By the law of conservation of
momentum, if the incident particle is deflected towards
the top, the screen will recoil towards the bottom and vice-versa. 
Einstein contended that by measuring the recoil momentum of screen 1, 
 it is in principle possible to determine which slit each particle passed
through. Successive light particles would eventually build up an
interference pattern.  Einstein argued that such an experiment would
show a violation of the principle of complementarity.

Bohr responded to Einstein's experiment by pointing out some subtleties
involved in obtaining the which-path information. According to Bohr,
to obtain  knowledge of  which slit the particle had passed
through, it was necessary to measure the movement of the screen to a
certain degree of accuracy. Any lesser degree of accuracy in the
measurement will fall short of providing the which-path information.
Consequently, screen 1 should be so sensitive that it should be treated
like a quantum object. In order that the recoiling screen gives a 
well-defined value of the momentum, its initial momentum should be
known to a good accuracy.
However, due to the uncertainty principle, there
will be a degree of uncertainty as to the position of the slit if the
momentum of the screen is well-defined. The
uncertainty in the position of the slits would lead to a superposition of
several slightly shifted patterns, sufficient to eliminate
the interference pattern \cite{recoil}.

Wooters and Zurek carried out a quantitative analysis of Bohr's argument,
assuming the recoiling slit to be constrained by a harmonic oscillator
potential \cite{wooters}. The experiment was realized in a very
interesting manner by the group of S. Haroche who has been awarded the
Nobel Prize in physics for 2012 \cite{haroche}. They were able to
set up a recoiling slit which could be tuned continuously from being
quantum-mechanical to being classical. Utter and Feagin realized the experiment
by using a trapped ion in place of the recoiling slit \cite{utter}.

\section{Theoretical Analysis}

\subsection{Which-path information and entanglement}

Bohr's reply to Einstein's recoiling slit experiment relies on the
assumption that the recoil experienced by the single slit will necessarily
disturb the state of the particle. This disturbance would  be just enough to wash out the interference. However, there
is a crucial aspect of getting which-path information, which is not part
of Bohr's reply. According to von Neumann, a quantum measurement consists
of two processes \cite{neumann}. The first one is a unitary evolution which
takes the product state of the system and detector to an entangled state,
which establishes correlation between states of the system and those
of the detector. For example, if the system is initially in a state
$\sum_{i=1}^n c_i|\psi_i\rangle$ and the detector is in a state $|d_0\rangle$,
the first process has the following effect:
\begin{equation}
 |d_0\rangle \sum_{i=1}^n c_i|\psi_i\rangle
\xrightarrow{\textrm{Unitary evolution}} \sum_{i=1}^n c_i |d_i\rangle |\psi_i\rangle .
\label{process1}
\end{equation}
The second process is essentially a non-unitary one which picks out
a single result (say) $|d_k\rangle |\psi_k\rangle$, from the superposition
on the right hand side of the above, with a probability
$|c_k|^2$. How such a superpostion of possibilities go over to a single
outcome in a measurement, is what constitutes the heart of the
{\em measurement problem}.
Here we will only be concerned with the first process. If we
apply the preceding argument to the case of Einstein's recoiling slit
experiment, there will be two orthogonal states of the particle,
$|\psi_1\rangle$ and $|\psi_2\rangle$, and two momentum states of the
recoiling slit, $|d_1\rangle$ and $|d_2\rangle$. There are two points to
be noted here.
\begin{enumerate}
\item[(a)] Two different momentum states of the recoiling slit will necessarily get
entangled with the states of the particle passing through the two slits.
\item[(b)] In principle it is possible to find an interaction which will
not affect the states of the particle $|\psi_1\rangle$ and $|\psi_2\rangle$,
but will only result in the detector states getting correlated with them.
\end{enumerate}
Point (a) is something which was not part of Bohr's reply. It can be easily
shown that this point alone is enough to rule out any interference pattern
for the particle.
The combined state of the recoiling slit and the particle,
on reaching the screen after passing through the
double-slit, will be of the form
\begin{equation}
\Psi_x = c_1|d_1\rangle\psi_1(x,t) + c_2|d_2\rangle\psi_2(x,t)
\end{equation}
The probability of finding the particle at a point $x$ on the screen is
given by 
\begin{eqnarray}
|\Psi_x|^2 &=& |c_1|^2\langle d_1|d_1\rangle|\psi_1(x)|^2
+ |c_2|^2\langle d_2|d_2\rangle|\psi_2(x)|^2 \nonumber\\
&+& c_1^*c_2\langle d_1|d_2\rangle\psi_1^*(x)\psi_2(x)
+ c_2^*c_1\langle d_2|d_1\rangle\psi_2^*(x)\psi_1(x)\nonumber\\
\end{eqnarray}
The last two terms in the above denote interference, and will vanish if
the two states of the recoiling slit, $|d_1\rangle$ and $|d_2\rangle$
are distinguishable, i.e., orthogonal to each other. Thus, the very fact
that which-path information is carried by the recoiling slit is enough to
rule out interference. One need not invoke the position-momentum uncertainty of
the recoiling slit. This will happen irrespective of the method one uses
to get which-path information in any other variant of this experiment.
The argument here is on the lines of the treatment of a particle going
through the double-slit interacting with a 1-bit detector by Scully,
Englert and Walther \cite{scully}.

If the inevitable entanglement in the measurement process, and its
implications had been recognized, Bohr could have provided a rebuttal
to Einstein without invoking the position-momentum uncertainty of the
recoiling slit.

\subsection{Path-distinguishability and fringe visibility}

Let us now look at the more interesting situation where the paths of the
particle through the two slits are only imperfectly distinguishable.
The following analysis is closely similar to that of Englert
where he derives the well-known Englert-Greenberger-Yasin duality relation
\cite{englert}. We define the distinguishability of the two paths by
\begin{equation}
{\mathcal D} = \sqrt{1 - |\langle d_1|d_2\rangle|^2},
\label{dist}
\end{equation}
where $|d_1\rangle$ and $|d_2\rangle$ are assumed to be normalized,
but not necessarily orthogonal to each other.
Clearly, for completely orthogonal $|d_1\rangle$ and $|d_2\rangle$,
${\mathcal D} = 1$, and for identical $|d_1\rangle$ and $|d_2\rangle$,
${\mathcal D} = 0$. If $|d_1\rangle$ and $|d_2\rangle$ are orthogonal
to each other, one can find an observable of the detector, for which
the two states can give two distinct eigenvalues. Measuring such an
observable, one can find out which of the two slits the particle
went through.

If the two states are not completely orthogonal, 
one can write the state $|d_2\rangle$ in terms of a component parallel
to $|d_1\rangle$ and a component orthogonal to it, 
\begin{equation}
|d_2\rangle = c_w|d_1\rangle+c_r|d_r\rangle,
\end{equation}
where $|d_r\rangle$ is a state orthogonal to $|d_1\rangle$, and
$|c_r|^2+|c_w|^2=1$.  Now, suppose one measures an observable which
gives different eigenvalues for $|d_1\rangle$ and
$|d_r\rangle$. If the particle goes through slit 1, the detector will
always give one particular outcome (right answer). However, if the particle
goes through slit 2, the detector will give a different value (right answer)
with probability $|c_r|^2$, but will give the same value associated with
slit 1 (wrong answer) with probability $|c_w|^2$. So,
one cannot distinguish the two paths with certainty if the $|d_1\rangle$
and $|d_2\rangle$ are not orthogonal. One can distinguish the two path
with a probability $|c_r|^2$. Using (\ref{dist}) one can verify that
$|c_r|^2 = {\mathcal D}^2$. Thus, in this case the probability with which
the two paths can be distinguished is equal to ${\mathcal D}^2$.
In general ${\mathcal D}^2$ can be considered to be the likelihood of
getting the correct which-way information.

Let us assume that a particle traveling along the z-direction passes
through a double-slit and also interacts with a which-path detector. We model the particle state as a Gaussian form with width $\epsilon$ when it strikes the slits. 
Through a process such as the one described by (\ref{process1}), the
detector states get correlated with the states of the particle emerging from
 the two slits. If this happens at time $t=0$, then we can write the combined state of the particle and the which-path
detector, in the form
\begin{equation}
\Psi(x,0) = A \left(|d_1\rangle e^{-{(x-d/2)^2\over 4\epsilon^2}}
+ |d_2\rangle e^{-{(x+d/2)^2\over 4\epsilon^2}}\right),
\label{entstate}
\end{equation}
where $A = {1\over\sqrt{2}}(2\pi\epsilon^2)^{-1/4}$. We have not given
different momenta to the two wave-packets which might result from a
``momentum back-action" of the which-way detector. We will show that 
just the correlation between the two wave-packets and the different states
of the recoiling slit is enough to destroy interference.
Here we do not expicitly consider the dynamics of the particle in
the z-direction. We just assume that the wave-packets are moving in the
forward direction with an average momentum $p_0=h/{\lambda}$, where 
${\lambda}$ is the d'Broglie wavelength of the particle.
Thus the distance $L$ travelled by the particle in a time $t_L$ is given
by $L = {h\over m{\lambda}}t_L$. This can be rewritten as
$\hbar t_L/m = {\lambda}L/2\pi$.

After a time $t$, the combined state of the particle and the detector evolves to
\begin{equation}
\Psi(x,t) = A_t \left(|d_1\rangle e^{-{(x-d/2)^2\over 4\epsilon^2+2i\hbar t/m}}
+ |d_2\rangle e^{-{(x+d/2)^2\over 4\epsilon^2+2i\hbar t/m}}\right),
\label{entstatet}
\end{equation}
where $A_t = {1\over\sqrt{2}}[\sqrt{2\pi}(\epsilon+i\hbar t/2m\epsilon)]^{-1/2}$.
The probability of finding the particle at position $x$ on the screen is
given by
\begin{eqnarray}
|\Psi(x,t)|^2 &=& |A_t|^2 \left(e^{-{(x-d/2)^2\over 2\sigma_t^2}}
+ e^{-{(x+d/2)^2\over 2\sigma_t^2}}\right)\nonumber\\
&+& |A_t|^2 \left(\langle d_1|d_2\rangle e^{-{x^2+d^2/4\over 2\sigma_t^2}}
e^{{ixd\hbar t/2m\epsilon^2\over 2\sigma_t^2}} \right.\nonumber\\
&& +\left. \langle d_2|d_1\rangle e^{-{x^2+d^2/4\over 2\sigma_t^2}}
e^{-{ixd\hbar t/2m\epsilon^2\over 2\sigma_t^2}}\right) ,
\end{eqnarray}
where $\sigma_t^2 = \epsilon^2 + (\hbar t/2m\epsilon)^2$. Writing
$\langle d_2|d_1\rangle$ as $|\langle d_2|d_1\rangle|e^{i\theta}$, and
putting $\hbar t/m = {\lambda}L/2\pi$, the above can be simplified to
\begin{eqnarray}
|\Psi(x,t)|^2 &=& 2|A_t|^2 e^{-{x^2+d^2/4\over 2\sigma_t^2}}
\cosh(x d /2\sigma_t^2)\times \nonumber\\
&&\left(1 + |\langle d_1|d_2\rangle| 
{\cos\left({{xd{\lambda}L/2\pi \over 4\epsilon^4+({\lambda}L/2\pi)^2}+\theta}\right)\over
\cosh(x d /2\sigma_t^2)} \right)
\label{pattern}
\end{eqnarray}
Eqn.(\ref{pattern}) represents an interference pattern with a fringe width
given by 
\begin{equation}
w = 2\pi\left({({\lambda}L/2\pi)^2 + 4\epsilon^4\over{\lambda}dL/2\pi}\right)
= {{\lambda}L\over d} + {16\pi^2\epsilon^4\over{\lambda}dL}.
\end{equation}
For $\epsilon^2 \ll {\lambda}L$ we get the familiar Young's double-slit
formula $w \approx {\lambda}L/d$.

Visibility of the interference pattern is conventionally defined as 
\begin{equation}
{\mathcal V} = {I_{\rm{max}} - I_{\rm{min}} \over I_{\rm{max}} + I_{\rm{min}} } ,
\end{equation}
where $I_{\rm{max}}$ and $I_{\rm{min}}$ represent the maximum and minimum intensity
in neighbouring fringes, respectively. In practice, fringe visibility will
depend on many things, including the width of the slits. For example, if
the width of the slits is very large, the fringes may not be visible at all.
Maxima and minima of (\ref{pattern}) will occur at points where the 
value of cosine is 1 and -1, respectively.
The two wave-packets emerge from two
narrow slits and quickly expand and overlap because of time evolution.
The width of a wave-packet, which was $\epsilon$ initially, is now
$\sigma_t = \sqrt{\epsilon^2 + (\hbar t/2m\epsilon)^2}$ at time $t$.
For sufficiently large $\sigma_t$, $\cosh(x d /2\sigma_t^2)$ can be
assumed to be $x$-independent over distances of the order of
fringe separation.  The visibility can then be written down as
\begin{equation}
{\mathcal V} = {|\langle d_1|d_2\rangle|\over \cosh(x d /2\sigma_t^2)}.
\end{equation}
Because $\cosh(y) \ge 1$, we get
\begin{equation}
{\mathcal V} \le |\langle d_1|d_2\rangle|.
\label{visiblity1}
\end{equation}
Using (\ref{dist}) the above equation gives a very important result
\begin{equation}
{\mathcal V}^2 + {\mathcal D}^2 \le 1.
\label{duality}
\end{equation}
This relation generalizes Bohr’s complementarity principle of mutual
exclusivity of wave and particles natures, to quantifying the extent
to which both these natures can be observed at the same time.
It sets a bound on the which-path 
distinguishability and the visibility of interference that one can
obtain in a single experiment. This inequality was derived earlier by
Greenberger and Yasin \cite{greenberger} and Englert \cite{englert} in
a more general context.

\subsection{Uncertainty and duality}

While it has been argued that the duality relation (\ref{duality}) is
independent of any kind of uncertainty relation \cite{greenberger,englert},
there is also another
view prevalent in the literature which holds that the process of which-way
detection introduces certain uncontrollable phases to the state of
the particle, which leads to loss of interference \cite{tan,storey}.
The uncertainty relation is believed to play a role in the latter.
Whether complementarity arises
out of correlations between the particle and a which-path detector or
from the uncertainty principle, has been a subject of some controversy
\cite{storey,englert2,wiseman,barad}.
Linked to this controversy is also the question whether the particle
receives any momentum kick from the recoiling slit, affecting its interference
pattern \cite{wiseman2,durr,unni}.
There have been various approaches to connect complementarity to 
uncertainty relations \cite{bjork,marzlin,huang,bosyk}. We explore
this issue in the light of the preceding discussion.

Let us suppose that a particle passing through a double-slit interacts with
a which-way detector, the recoiling slit in our case. Let us suppose that
corresponding to particle passing through slits 1 and 2, the recoiling slit
acquires two distinct momentum states. If this is true, we can always find an
observable $\op{P}$ which will give eigenvalues (1,-1) corresponding to the
particle passing through slits 1 and 2, $\op{P}|p_1\rangle=|p_1\rangle$ and
$\op{P}|p_2\rangle=-|p_2\rangle$.

In a non-ideal situation,  the recoiling slit may have  only 
partial which-way information. Then
the states that actually get correlated with the particle paths in
(\ref{entstate}) could be written as
\begin{equation}
|d_1\rangle = c_1|p_1\rangle +c_2|p_2\rangle,~~
|d_2\rangle = c_2^*|p_1\rangle +c_1^*|p_2\rangle,
\end{equation}
With the provision $|c_1|^2+|c_2|^2=1$, $|d_1\rangle$ and $|d_2\rangle$ are 
normalized but not necessarily orthogonal. The ideal situation would
correspond to $|c_1|=1,~c_2=0$ or vice-versa, where the detector states
would carry full which-way information. For the case $|c_1|=|c_2|=1/\sqrt{2}$,
the detector states would carry no which-way information. Thus, the above
form of $|d_1\rangle,~|d_2\rangle$ covers all possibilities of mutual
overlap.

The square of uncertainty in
$\op{P}$ with any of these two states, is given by
\begin{equation}
\Delta P^2 = \langle\op{P}^2\rangle - \langle\op{P}\rangle^2
 = 4|c_1|^2|c_2|^2
\end{equation}
Note that distinguishablity, as defined by (\ref{dist}), now has the form
\begin{equation}
\mathcal{D}^2 = 1 - 4 |c_1|^2|c_2|^2 .
\end{equation}
This implies that
\begin{equation}
\mathcal{D}^2 = 1 - \Delta P^2
\label{dist2}
\end{equation}
So, for distinguishablity to be 1, $\Delta P$ should be zero.

If one does not wish to talk in the language of correlations between the
particle and the detector carrying which-way information,
and just wants to look at interference build-up from individual particles
registering on the screen, one has to take into account
the change in relative phase of the
amplitudes of particle passing through the two slits because of
interaction with the which-way detector \cite{unni}. This was the approach
taken by Bohr in replying to Einstein's recoiling slit experiment.
If the particle paths are correlated to $|p_1\rangle$ and $|p_2\rangle$,
so that the combined state is 
$\Psi(x) = \psi_1(x)|p_1\rangle + \psi_2(x)|p_2\rangle$, the
particle state corresponding to a detector state
$(|p_1\rangle+|p_2\rangle)/\sqrt{2}$ is $[\psi_1(x) + \psi_2(x)]/\sqrt{2}$,
that corresponding to the detector state 
$(|p_1\rangle-|p_2\rangle)/\sqrt{2}$ is $[\psi_1(x) - \psi_2(x)]/\sqrt{2}$.

Thus, there exists another observable of the recoiling slit, $\op{Q}$,
such that
$\op{Q}|q_1\rangle = |q_1\rangle$, $\op{Q}|q_2\rangle = -|q_2\rangle$,
and $|q_1\rangle = (|p_1\rangle+|p_2\rangle)/\sqrt{2}$,
$|q_2\rangle = (|p_1\rangle-|p_2\rangle)/\sqrt{2}$.
Observables $\op{Q}$ and $\op{P}$ obviously do not commute, and both
can be represented by Pauli spin operators. For instance, if
$\op{P}=\op{\sigma_z}$, then $\op{Q}=\op{\sigma_x}$.

\begin{figure}[h!]
\centering
\includegraphics[width=3.5 in]{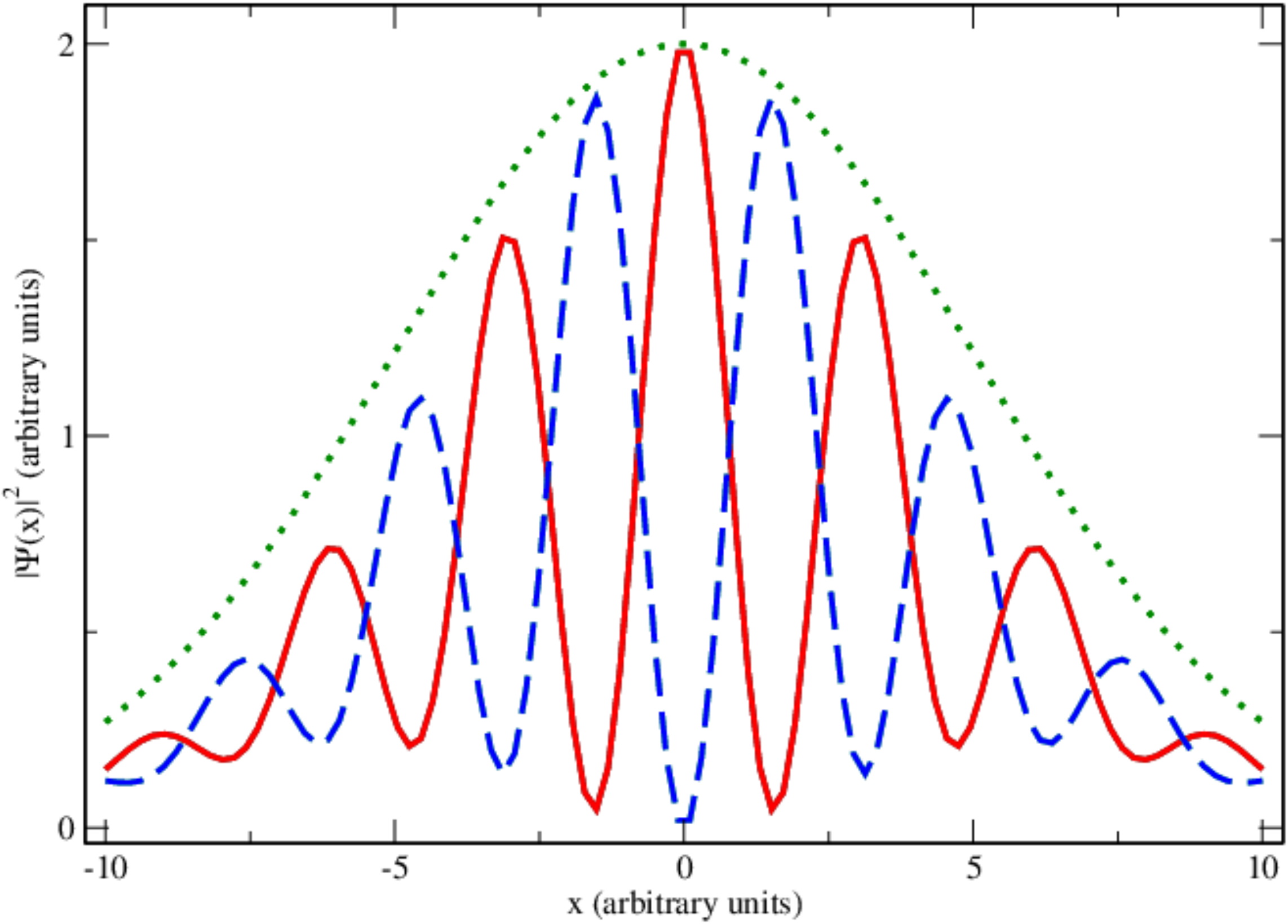}
\caption{If one correlates the detected particles with measurement results
on $\op{Q}$, one gets two complementary patterns corresponding to
$|q_1\rangle$ (solid line) and $|q_2\rangle$ (dashed line). Without any
correlation, there is no intference pattern (dotted line).  }
\label{twopatterns}
\end{figure}

In general, the state of a particle passing through the double-slit and
interacting with the recoiling slit, can be written in terms of the
eigenstates of operator $\op{Q}$, as follows
\begin{equation}
\Psi(x) = {c_1\over\sqrt{2}}[\psi_1(x)+\psi_2(x)]|q_1\rangle +
{c_2\over\sqrt{2}}[\psi_1(x)+\psi_2(x)]|q_2\rangle
\label{entstate2}
\end{equation}
with $|c_1|^2 +|c_2|^2=1$. For $|c_1|=|c_2|=1/\sqrt{2}$, and $\psi_1,~\psi_2$
having the form represented in (\ref{entstatet}), if one correlates the
probability of finding the particle on the screen with the measured eigenstate
of $\op{Q}$, (\ref{entstate2}) yields two shifted interference patterns
(see FIG. \ref{twopatterns}). But together, the two patterns kill each other. This is
an example of quantum eraser \cite{eraser}, where correlating the detected
particle with certain states of the which-way detector can ``erase" the
which-way information and the interference pattern can be observed.

For any $c_1,~c_2$, the probability of finding the particle on the
screen is given by
\begin{eqnarray}
|\Psi(x)|^2 &=& {1\over 2}\left(|\psi_1(x)|^2+|\psi_2(x)|^2\right)\nonumber\\
&&+ {|c_1|^2-|c_2|^2\over 2}\left(\psi_1^*(x)\psi_2(x)+\psi_2^*(x)\psi_1(x)\right).\nonumber\\
\end{eqnarray}
Using the earlier analysis, in particular (\ref{visiblity1}), we can
immediately write
\begin{equation}
{\mathcal V}^2 \le (|c_1|^2-|c_2|^2)^2.
\end{equation}
The uncertainty in the observable $\op{Q}$, in the state (\ref{entstate2}),
can be evaluated to yield
\begin{equation}
\Delta Q^2 = 1 - (|c_1|^2-|c_2|^2)^2.
\end{equation}
While doing so, one has to make use of the fact that $\psi_1(x)$ and
$\psi_2(x)$ are orthonormal states \cite{ndh-tq}. The two equations above yield
\begin{equation}
{\mathcal V}^2 \le 1 - \Delta Q^2.
\label{visiblity2}
\end{equation}
Using (\ref{dist2}) and (\ref{visiblity2}) we find
\begin{equation}
{\mathcal D}^2 + {\mathcal V}^2 \le 2 - [ \Delta P^2 + \Delta Q^2].
\label{duality2}
\end{equation}
For any two-level system, adequately described by Pauli spin matrices,
any two spin-components satisfy the following sum uncertainty relation
\cite{hofmann}
\begin{equation}
\Delta\sigma_1^2 + \Delta\sigma_2^2 \ge 1,
\end{equation}
which in turn implies that $\Delta P^2 + \Delta Q^2 \ge 1$. Using this
result, (\ref{duality2}) reduces to
\begin{equation}
{\mathcal D}^2 + {\mathcal V}^2 \le 1.
\label{duality3}
\end{equation}
Thus we find that the Englert-Greenberger-Yasin duality relation also emerges
as a consequence of the sum uncertainty relation for certain observables
of the recoiling slit.

At this stage it might be useful to make connection between the preceding
analysis and Bohr's reply to Einstein. Bohr had argued that a
fixed position of the recoiling slit would correspond to a sharp interference.
Different fixed positions of the recoiling slit would correspond
to slightly shifted interference patterns. Our analysis shows that two
distinct values of $Q$ lead to two sharp, but mutually shifted, interference
patterns (see FIG. \ref{twopatterns}). In our analysis, for an accurate
which-way information, one needs an eigenvalue of $P$ , which will result in a
superposition of two values of $Q$, and consequently to a superposition of
two shifted interference patterns. In Bohr's argument, a distinct value
of momentum would lead to a superposition of different positions of
the recoiling slit, and a superposition of many shifted interference patterns,
and hence, loss of interference. Our $P$ and $Q$ are analogous to
the momentum and position of the recoiling slit, respectively, in Bohr's
argument. Thus, the preceding calculation may be viewed as a more
quantitative analysis of Bohr's estimate.  The Englert-Greenberger-Yasin
relation emerges as a more general statement of the Heisenberg uncertainty
relation as invoked by Bohr in the context of Einstein’s recoiling slit
experiment.

\section{Discussion}

The preceding analysis, although applicable to
Einstein's recoiling slit experiment, is fairly general. The detector
states $|d_1\rangle$
and $|d_2\rangle$ may correspond to states of any other kind of which-way
detector.  In the analysis of Wooters and Zurek \cite{wooters}, the
recoiling slit was modelled as a harmonic oscillator in its ground state,
which is a Gaussian state with zero average momentum. The states $|d_1\rangle$
and $|d_2\rangle$ here correspond to Gaussian states with oppositely shifted
average momentum.  The distinguishability of the two states, as defined by
(\ref{dist}), will put a bound on the visibility of the interference pattern,
according to (\ref{duality}).

From the preceding analysis we see that interefrence visiblity ${\mathcal V}$
can be 1 only when $\Delta Q$ is 0. Also, the which-path distinguishability
${\mathcal D}$ can be only if $\Delta P$ is 0. Because $\op{P},~\op{Q}$ do not
commute, $\Delta P,~\Delta Q$ cannot both be zero at the same time.
This can also be assumed to be a fundamental reason enforcing complementarity.

Bohr's argument of the position uncertainty of the recoiling slit did
rule out simultaneous observation of interference and obtaining which-path
information. This led many to believe that Bohr's complementarity principle
was in fact, a tacit restatement of the position-momentum uncertainty relation.
However, the sum uncertainty relation for observables $\op{P}$ and $\op{Q}$,
introduced here, puts a tighter bound on fringe visibility and which-way
information. It actually leads to the very fundamental
Englert-Greenberger-Yasin duality relation. So, the sum uncertainty of
certain two-state observables seems to be enforcing complementarity in
a more fundamental way than the Heisenberg uncertainty relation.

From the analysis of section IIIB we have seen that the
Englert-Greenberger-Yasin duality relation also comes out from the
correlation between the particle paths and which-path detector states.
On the other hand, if the particle paths get correlated to certain orthogonal
detector states, one can {\em always} find two observables $\op{P}$ and
$\op{Q}$, whose sum-uncertainty relation will be a quantitative
statement of complementarity.
Thus we see that the mutual exclusivity of wave
and particle nature emerges as a consequence of quantum correlation of
the particle with the which-way detector states, and also from the sum
uncertainty relation of certain observables of the which-way detector.
This indicates that uncertainty relations are as much an inherent part
of the which-way detection process, as are the quantum correlations.
So, quantum correlations and quantum uncertainty relation are two
alternate ways of looking at the same phenomenon. Both lead to the
fundamental Englert-Greenberger-Yasin duality relation.

Lastly, we point out that there has been a prevailing view that Bohr's
reply to Einstein implied that the particle receives momentum kicks due to
its interaction with the detector, and that enforces complementarity.
We emphasize that Bohr never talked about any momentum back-action
on the particle from the recoiling slit. He only said that the particles
originating from a particular
position (position of the recoiling slit) will lead to a particular position
of the interference pattern.
A shifted position of the recoiling slit, will lead to a shifted pattern.
If there is an uncertainty in the position of the recoiling slit, it will
lead to an uncertainty in the location of the fringes, and hence washing
out of interference. In our analysis of section IIIC, the two eigenstates
of $\op{Q}$ lead to two different locations of the interference pattern
(see FIG. 2). So, $\op{Q}$ in our analysis plays the role of position of
the recoiling slit in Bohr's argument. Here interference loss is due to
different relatives phases associated with the two particle paths,
corresponding to different eigenstates of $\op{Q}$. Any 
momentum back-action on the particle is an additional baggage, not
essential to explaining the loss of interference.

\begin{acknowledgments}
This work was concieved during the {\em International Workshop on Quantum
Information-2012}, at Allahbad. The authors thank the organisers for
providing a platform for exchange of ideas.
\end{acknowledgments}

\end{document}